\documentclass[conference]{IEEEtran}
\IEEEoverridecommandlockouts
\usepackage{cite}
\usepackage{amsmath,amssymb,amsfonts}
\usepackage{algorithmicx}
\usepackage{graphicx}
\usepackage{textcomp}
\usepackage{xcolor}
\usepackage{lmodern}
\usepackage[utf8]{inputenc}
\usepackage[T1]{fontenc}
\usepackage{CJKutf8}

\def\BibTeX{{\rm B\kern-.05em{\sc i\kern-.025em b}\kern-.08em
    T\kern-.1667em\lower.7ex\hbox{E}\kern-.125emX}}
\begin{document}
%
\title{An Automatic Debugging Tool of Instruction-Driven Multicore Systems with Synchronization Points\\
{\footnotesize }
\thanks{}%
}%
\author{
\IEEEauthorblockN{1\textsuperscript{st} Yuzhe Luo}
\IEEEauthorblockA{\textit{State Key Laboratory of Computer Architecture, Institute of Computing Technology, CAS} \\
\textit{University of Chinese Academy of Sciences}\\
Beijing, China \\
luoyuzhe16@mails.ucas.ac.cn
}
\and
\IEEEauthorblockN{2\textsuperscript{nd} Xin Yu}
\IEEEauthorblockA{\textit{Institute of Computing Technology,Chinese Academy of Sciences} \\
Beijing, China \\
yuxin@ict.ac.cn
}
}%

\maketitle

\begin{abstract}
Tracing back the instruction execution sequence to debug a multicore system can be very time-consuming because the relationships of the instructions can be very complex. For instructions that cannot be checked by the environment immediately after their executions, the errors they triggered can propagate through the instruction execution sequence. Our task is to find the error-triggered instructions automatically. This paper presents an automatic debugging tool that can leverage the synchronization points in the instruction execution sequences of the multicore system being verified to locate the instruction which results in simulation error automatically. To evaluate the performance of the debugging tool, we analyze the complexity of the algorithms and count the number of instructions executed to locate the aimed instruction.
\end{abstract}


\section{Introduction}
As the Design Under Test(DUT) becomes more and more complex, verification becomes more time-consuming because the verification engineer has to analyze tedious logs and waveforms to locate the bugs in RTL codes. Take checking data races as an example, if data races happen in the verification of complex multicore system with long executed instruction sequences the UVM scoreboard may find that the values in RAMs are wrong. Since the location where the wrong values in RAMs can be read and written many times, the wrong values can be propagated to many other places in RAMs or register files so that it will be very hard to find the instructions that cause the data races. To locate the instruction which causes errors quickly, we design the automatically debugging tool.
We define the instruction for which the environment reports error as the error-reported instruction. We define the instruction which triggers a hardware bug as the error-triggered instruction. Not all instructions can be checked immediately because (1) Checking all the instructions immediately can increase the complexity of the environment significantly, (2) the design of the DUT can make it hard to get the result immediately after its execution, and (3) some instructions do not output results such as the jump instruction[1].
We define the instructions that are checked by the scoreboard immediately after their executions as the immediately checked instructions. We define the instructions that cannot be checked immediately by the scoreboard as the lazily checked instructions. When the immediately checked instructions and the lazily checked instructions are mixed, an error-reported instruction may not be an error-triggered instruction because the error-triggered instruction may not be checked immediately after its execution so that its wrong results influence the instructions executed after it. Tracing back the instruction sequences with complex dependency relationships manually for a multicore system can cost a lot of time.
In this paper, we focus on debugging the multicore systems which is popular nowadays and propose an automatic debugging tool that leverages the synchronization points to help locating the error-triggered instructions. The current works about multicore systems concentrate on accelerating verification by improving the technique of random test generation[2],[4],[5],[7] and verifying cache coherency[2],[3],[4]. Also, some works are related to the verification of timings[6] and speed debugging the DUT[8]. However, to the best of our knowledge, there are few works related to debugging a multicore system automatically in the verification field. We can learn from the tracing and snapshots technique used in debugging software[9]. Besides, many classic works related to memory consistency verification have been published.[10] builds a memory consistency verification tool that can verify several memory consistency models quickly. To reduce the complexity of verifying the memory consistency, some microarchitecture dependent methods which leverage extra observations in the design have been proposed[11][12][13] while other microarchitecture independent methods which propose polynomial-time algorithms may not be complete[14][15][16][17]. We use so-called synchronization regions to accelerate finding the instructions cause data races.
In section II we introduce the workflow of the automatically debugging tool and proposes a tree-search algorithm to locate the error-triggered instruction. In section III we explain the implementation details of the automatically debugging tool. In section IV we evaluate the performance of the automatic debugging tool.

\section{THE WORKFLOW OF THE DEBUGGING TOOL}
In this section, we introduce the workflow of the debugging tool. To locate the error-triggered instruction, the debugging tool firstly divides the instruction sequences into synchronization regions. Then it builds the dependency graph of the synchronization regions. After that, it builds instruction dependencies trees for incorrect machine state snapshots and locates the synchronization region where the error-triggered instruction exists. Finally, it uses the instruction dependency tree of the located synchronization region to locate the error-triggered instruction. We will introduce the four steps in detail with examples below.
\subsection{Definitions}
Firstly, we make some definitions clear. A synchronization point is an instruction that guarantees the instructions before it finishes their executions before the executions of the instruction following it. Inter-core synchronization points synchronize the instruction execution of different cores. That is to say, the instruction execution meets the inter-core synchronization points earlier will wait for the instruction execution of other cores which also contains the inter-core synchronization point to meet the synchronization point. A synchronization region is the instruction sequence between two adjacent synchronization points.
A machine state snapshot is a set of files storing the values in RAMS and register files of DUT. The files are usually organized in a tree structure which is similar to the organization of the DUT. A model state snapshot is also a set of files that store the values provided by the reference model to be compared to the corresponding values in the machine state snapshot.
An instruction dependency tree can be generated from an error-reported instruction or a machine state snapshot storing wrong values. In an instruction dependency tree, every node stores a set of instructions that store data to the same location. A location can be an address interval of RAMs or a register.  The location which has been accessed by all the instructions stored in the node is also stored in the node. The address intervals or registers stored in the children of a node are those where the operators of the instructions stored in the parent node are stored. To generate an instruction dependency tree, we first find the location where the error reported instruction output its result or the wrong values in the machine state snapshot are stored. Two instructions are data-dependent if one uses another instruction's output values as its operators. We extend the instruction dependency tree according to the instruction data dependency relationships.
\begin{figure}[htbp]
\centerline{\includegraphics{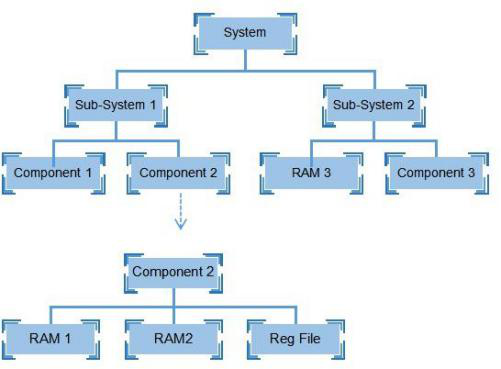}}
\caption{an example of machine state snapshot}
\label{fig}
\end{figure}

\begin{figure}[htbp]
\centerline{\includegraphics{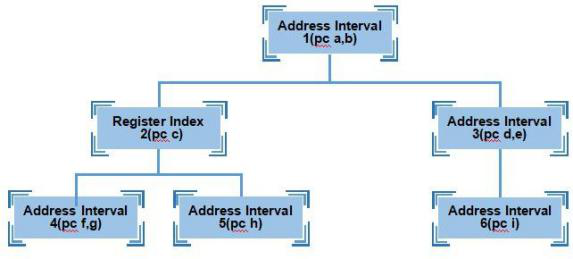}}
\caption{an instruction dependency tree}
\label{fig}
\end{figure}
\subsection{Dividing the instruction sequences into synchronization regions}
For a multicore system, each core can execute its instruction sequences. The cores can cooperate and accomplish a computation task together. Suppose there are several cores in the multicore system. Cores can share some RAMS and register files. Also, they have their own RAMs and registers. Synchronization points are used to synchronize different cores or synchronize the parallel instruction executions in a single core. Here we firstly consider the inter-core synchronization. We regard the initial state before running instructions and the final state after running the instruction sequence as synchronization points. We divide the instruction sequences into synchronization regions as below.
\begin{figure}[htbp]
\centerline{\includegraphics[width=10cm]{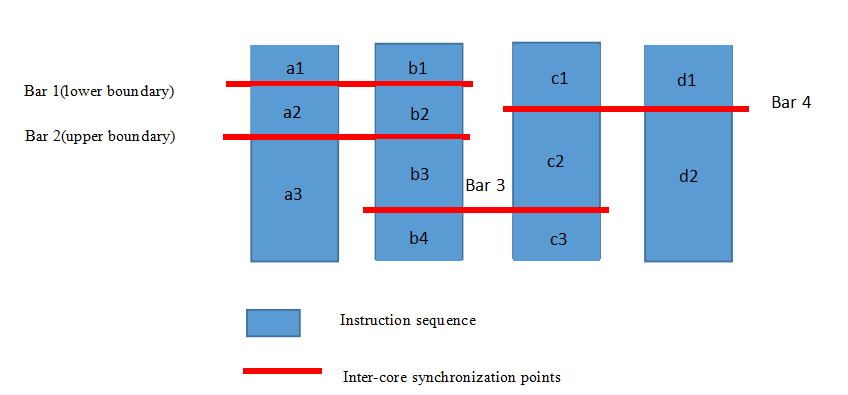}}
\caption{Dividing instruction sequence into synchronization regions}
\label{fig}
\end{figure}
The instruction sequences are divided by the inter-core synchronization points which are represented by the red bars. The execution of instruction sequences ends with the same bar will be synchronized. For example, if the execution of instructions in synchronization region a1 meets the bar before that in b1, then the execution flow of  a1 will wait for that of b1 to reach the bar.
\subsection{Building the dependency graph of the synchronization regions}
After dividing the instruction sequences into synchronization regions, we build a dependency graph with them. The dependency graph is a directed graph and a synchronization region at the head of the arrow is executed before the synchronization region at the tail of the arrow. If there is a path between two synchronization regions, then the synchronization regions are executed sequentially in order. If a synchronization region is not reachable from another synchronization region, then the two regions can be executed in parallel. The dependency graph building for Figure 3 is shown below.
\begin{figure}[htbp]
\centerline{\includegraphics[width=9cm]{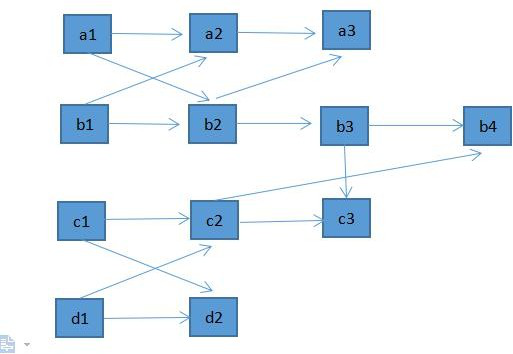}}
\caption{Dividing instruction sequence into synchronization regions}
\label{fig}
\end{figure}
To judge whether two synchronization regions can be executed in parallel quickly, we build linked lists for the dependency graph. A linked list stores all the synchronization regions that can be executed in parallel with the synchronization region in the head. For example, the linked list built for a1 is shown in Figure 5. The linked list consists of all the nodes that cannot be reachable from a1.
\begin{figure}[htbp]
\centerline{\includegraphics[width=9cm]{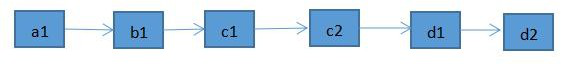}}
\caption{an example of a linked list with a1 as the head}
\label{fig}
\end{figure}
\subsection{Locating the synchronization region where the error-triggered instruction exists}
Suppose the reference model is golden. Before starting the simulation of DUT, the tool uses the reference model to generate the model state snapshots for all the inter-core synchronization points. After that, the tool starts the simulation process and compares the machine state snapshot with the corresponding model state snapshot when meeting an inter-core synchronization point. We call the synchronization region with the met synchronization point as the upper boundary as the current synchronization region and the synchronization regions that can be executed parallel with the current synchronization region as the parallel synchronization regions.
Using the linked list generated in the last step, the parallel synchronization regions are easy to be found. If the machine state snapshot is the same as the model state snapshot the instruction execution of the DUT will continue. Otherwise, for each address interval or register with a different value or the error-reported instruction met in the current synchronization region we build an instruction dependency tree with the address interval or the index of the register as the root. Then we extend the instruction dependency tree according to the instruction data dependency.
Each instruction dependency tree only consists of instructions in the current synchronization region or the parallel synchronization regions. If the extended branch meets an instruction from a synchronization region which is a predecessor of the current synchronization region, the extension of the branch stops. Because the machine state snapshots of the predecessors have been checked. After that, we search the dependency tree to make sure there are no data races. This is done by judge whether the instructions in a node and its children are from the same synchronization region. If it is not, it means that two instructions from different synchronization regions that can be run in parallel have data dependency. If a data race is checked,  the process stops and returns the instructions with the data race. If no data race is checked, then the tool checks whether the different values are all from instructions in the parallel synchronization regions. If it is, the simulation process continues because the parallel execution of instructions in the reference model and DUT can be different. Otherwise to the tool goes to step 4 and locates the error-triggered instruction exactly.
For example, in figure 3 if there is an instruction in synchronization region b2 and an instruction in c1 have data races, the two instructions must be stored in one node if they have write-after-write conflicts or in a parent node and one of its children separately if they have write-after-read conflicts or read-after-write conflicts. If there is an error-triggered instruction in b2 and there are no data races, the error-triggered instruction will store some wrong values in registers or RAMs which will be part of the machine state snapshot generated at bar 2 or be the operators of an error-reported instruction in the synchronization region. An instruction dependency tree can be built with the machine state snapshot or the error-triggered instruction. Then the debugging process will go to step
4. The machine state snapshots and the model state snapshots generated in this step will be stored in a database. The flowchart of the process above is shown below.
\subsection{Locating the error-triggered instruction exactly}
In this step, the debugging tool uses the data dependency tree generated last step to locate the
error-triggered instruction exactly. Recall that there are no data races for the instructions executed so far, hence there must be a data dependency tree where all the instructions in its nodes are in the current synchronization region. Now the problem is transferred to locating the error-triggered instruction in a single-core case which we have solved in[1]. We explain our idea by an example here.
If there is an error-triggered instruction in synchronization region b2 and we have built an instruction dependency tree which consists of the error-triggered instruction, we will use the machine state snapshot at the lower bound bar 1 to resume the simulation process for the instructions in the synchronization region b2 and use the reference model to execute the same instruction sequence. Each instruction in the nodes will be set as the finish instruction of the executed instruction sequence and the tool compares the machine state snapshot and the model state snapshot after executing the instruction sequence. If the machine state snapshots for all the instruction in a node is correct, the branch with the node as the root will be cut in the dependency tree. Otherwise, the process continues for the children of the root. The algorithm returns the instruction with the smallest pc value which generates wrong machine state snapshot. The algorithm is shown below.
\begin{figure}[htbp]
\centerline{\includegraphics[width=9cm]{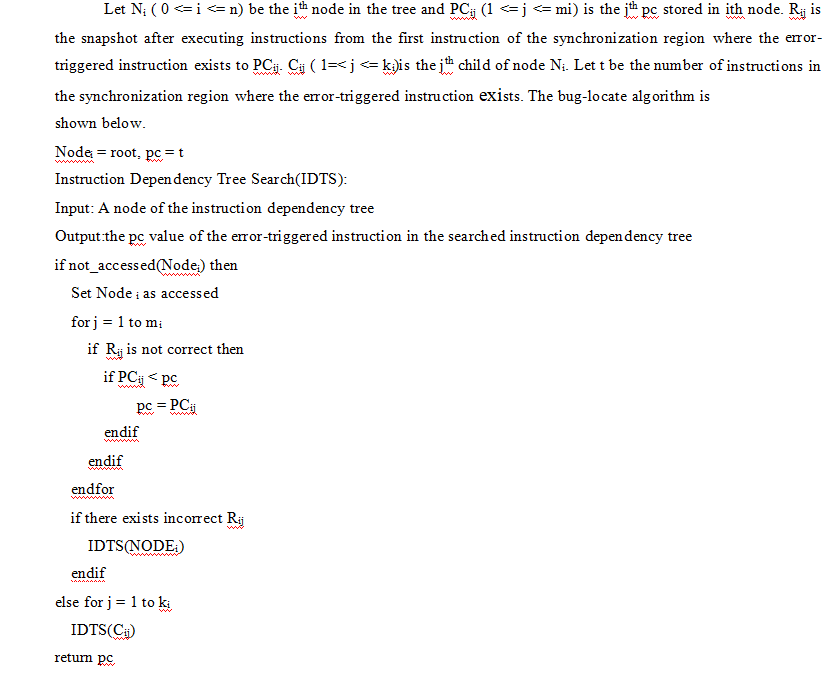}}
\end{figure}
\begin{figure}[htbp]
\centerline{\includegraphics[width=9cm]{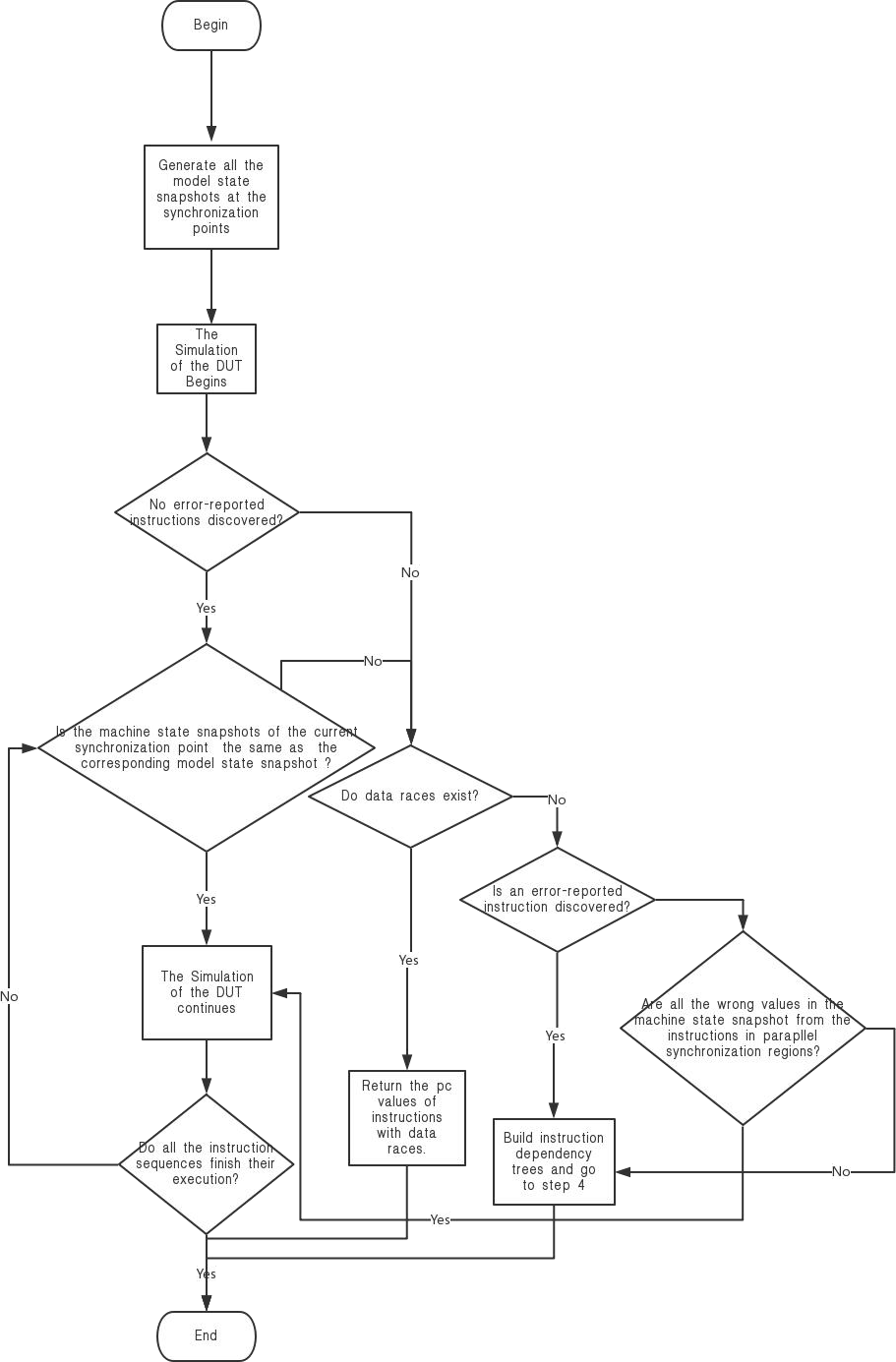}}
\caption{The flowchart for step 3}
\label{fig}
\end{figure}

\section{THE IMPLEMENTATION DETAILS OF THE DEBUGGING TOOL}
There are four major parts of the debugging tool,1) the analyzer, 2) the reference model, 3) the git-like data base and 4) the scoreboard. We Shown the architecture of the debugging tool in figure 7.
\begin{figure}[htbp]
\centerline{\includegraphics[width=9cm]{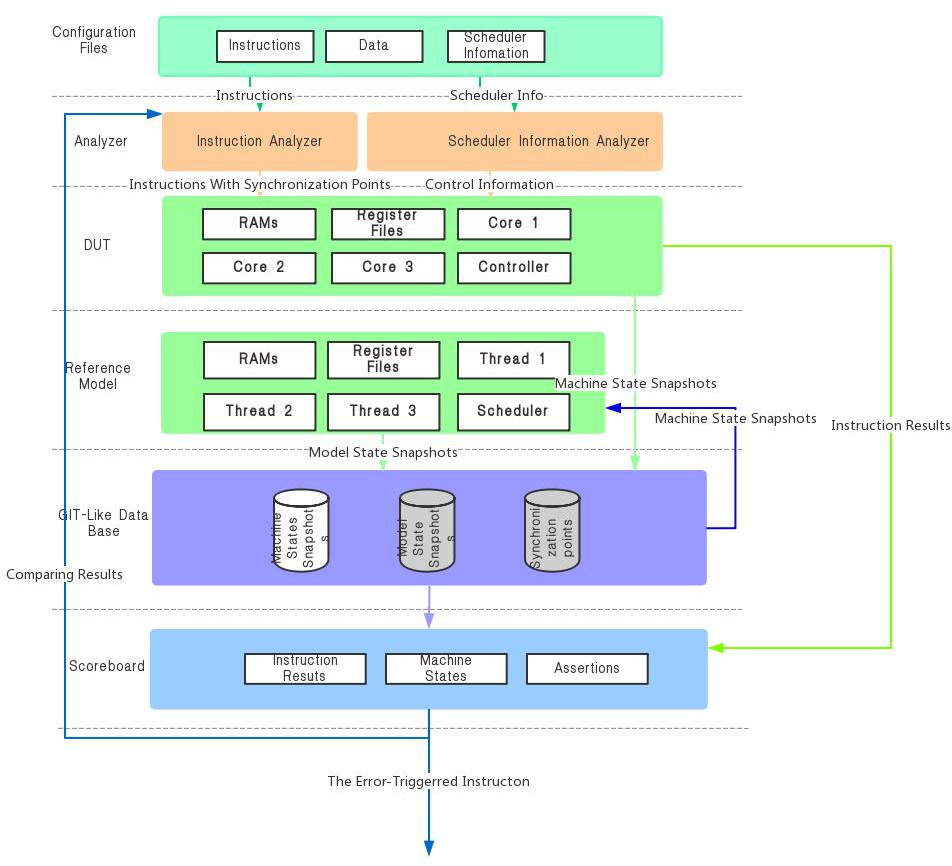}}
\caption{the architecture of the debugging tool}
\label{fig}
\end{figure}
\subsection{The analyzer}\label{AA}
The analyzer is made up of the instruction analyzer and the scheduler information analyzer. The instruction analyzer scans the instruction sequences and divide them into synchronization regions. Generally the debugging tool uses the synchronization point in the original instruction sequences to help locating the error-triggered instruction. The instruction analyzer can also insert synchronization points into the instruction sequences to reduce the length of the synchronization regions according to user-defined rules as long as the hardware bug can be triggered. Also the instruction analyzer can generate the instruction dependency tress according to the data dependency between the instructions and the comparing results of the instruction execution or the snapshots. The scheduler information analyzer analyzes the scheduler information and generate configuration files for the DUT and the reference model.

\subsection{The Git-like Database}
The database is used to store the machine state snapshots and model state snapshots. Since the snapshots are stored in tree-like directories which can be stored in the git file system. We leverage the operation used in the local git file system to realize the data base. We use git-add and git-commit to store the snapshots in the local git file system. Notice that the git-commit command will preserve all the historical data which can be fetched by the git-checkout command. Also, we develop two branches to store the machine state snapshots and model state snapshots separately. We use the git-diff command to differentiate between the machine state snapshots and the model state snapshots. Each snapshot generated by the DUT or the reference model will be stored in the database and the state snapshot of the same synchronization point will be updated if a new version is generated.

\subsection{The Scoreboard}
The scoreboard is generally a UVM scoreboard. It checks the correctness of results of the immediately checked instructions and supports the UVM assertions. In the debugging tool, the scoreboard compares the machine state snapshot with the model state snapshots of the same synchronization point. It can find the exact address interval of RAMs or the register where the different value exits and return the address interval or the indices of the registers to the instruction parser to build an instruction dependency tree. Also, the environment will judge whether to continue to run the DUT simulation according to the comparing results. If the machine state snapshot is different from the corresponding model state snapshot, we can imply that some instructions executed before the synchronization point output wrong results(including data races) or the execution order of the instruction sequence is in a wrong manner. The reason is that for computation and IO instructions, their results will finally be stored in RAMs or registers. For control instructions like jump instructions, if they change the instruction execution order in a wrong manner the value stored in RAMs and registers will be different from the corresponding value provided by the reference model.

\section{EVALUATIONS AND EXPERIMENTS}
In this section we compute the time complexity and space complexity of the whole process step by step. Let s be the number of inter-core synchronization points. Recall that we regard the initial state before executing instructions and the finial state after executing all the instructions as synchronization points. Let n be the total number of instructions in the instruction sequences. In the first step we just search through the instruction sequences and find all the inter-core synchronization points. The time complexity of this step is O(n). Since we need to divide the instruction sequences into synchronization regions and store their locations, the space complexity is O(s).
In the second step, to find all the synchronization regions that are reachable from a given synchronization region in the dependency graph in the worst case we must search all the remaining synchronization regions. Once we have done this for a synchronization region we can drop it in the following search process. The time complexity of the search process is O(s$^{2}$).As for the space complexity, notice that the number of nodes and pointers used in the graph is no more than s. And the linked lists store all the synchronization regions that can be executed in parallel and this is equal to add two edges with opposite directions between the nodes where there are no edges in the dependency graph because for two nodes the linked list with one node as the head will store the other and vice versa. Considering the number of pointers in the worse case the space complexity is O(s$^{2}$) while the number of nodes can be reduced to O(s) by reusing the nodes in the dependency graph.
In step 3 we need to execute all the instructions in the instruction sequences and store snapshots for every synchronization points hence the time complexity is O(n+s) and space complexity is O(s).
In the step 4, in the worst case the searched synchronization region storing most of the instructions in the instruction sequences and the instruction dependency tree can store most of the instructions in the synchronization region hence the time complexity is O(n$^{2}$) and the space complexity is O(n).To summarize, the time complexity of the whole process is O(n$^{2}$ + s$^{2}$).
But in practice, because in steps 3 and 4 we need to run the simulation which is much more time-consuming than running the C++ program, most of the time is consumed by step 3 and 4. In addition, we only consider the worst case above but the average case can be much quicker. For example, the number of instructions a synchronization region consists of can be much less than n hence on average the time complexity of step 4 will be much lower than the upper bound. Also, we do not need to execute all the instructions if an error is triggered early.
To evaluate the performance of the automatically debugging tool, we record the total number of instructions executed by the tool to locate the error-triggered instruction. We only count the instructions that must be executed to get the snapshot of the synchronization region where the error-triggered instruction exists. The instructions are those in the synchronization regions from which the synchronization region where the error-triggered instruction exist or in the synchronization regions which have the same upper boundary as the synchronization region where the error-triggered instruction exist. The multicore systems in the experiment have 5, 10, 20 or 30 cores. In each trial, we execute 500, 1000 or 1500 instructions with each core. We place the error-triggered instruction randomly in the instruction sequences in each trial and have 30 trials for each configuration. Finally, we calculate the average number of executed instructions for the 30 trials of each configuration. The experiment results are shown in the following form.
From the experiment results, we observe that the average number of instructions executed to locate the error-triggered instruction is much less than that estimated for the worst case. This is because of the length of a
The synchronization region is usually much less than the length of the instruction sequence of each core. Hence in step 4 we only need to build an instruction dependency tree for the short synchronization region where the error-triggered instruction exists. Also, if the error-triggered instruction is found early, the number of executed instructions will be reduced. Using the automatically bug-locating tool, we locate bugs automatically rather than debug with logs, waveform and repeat the whole simulation process. For complex multicore system and long instruction sequences, the verification engineers may spend several days for locating the error-triggered instruction. With this debugging tool, the error-triggered instruction can be located automatically in several hours.
\begin{table}[htbp]
\caption{Experiment Results}
\begin{center}
\begin{tabular}{|c|c|c|c|c|}
\hline
\textbf{Instruction}&\multicolumn{4}{|c|}{\textbf{Core Number}} \\
\cline{2-5}
\textbf{Number} & \textbf{\textit{5}}& \textbf{\textit{10}}& \textbf{\textit{20}}& \textbf{\textit{30}} \\
\hline
500& 1340& 2401& 5896 &6973  \\
\hline
\hline
1000& 2463& 4507 & 9769 & 13877 \\
\hline
\hline
1500& 3594& 7740 & 12124 & 22923 \\
\hline
\end{tabular}
\label{tab1}
\end{center}
\end{table}

\section{CONCLUSION}
In this paper, we introduce an automatic debugging tool for instruction-driven components and systems. The debugging tool leverages the synchronization points in the instruction sequence to observe the machine state of the DUT and find error in the simulation as early as possible. Also, it can locate data races and the error-triggered instruction automatically. As a result, it reduces the time of debugging dramatically and release verification engineers from tedious logs and waveform.

\section*{Acknowledgment}

\end{document}